\documentclass[aps,prl,12pt,showkeys]{revtex4}

\usepackage{amsmath}
\usepackage{amssymb}
\usepackage{times}
\usepackage{mathptmx} 

\begin{document}
\title{Deterministic cloning of an unknown Bell state}
\author{J. Orlin Grabbe}
\email[Email: ]{quantum@orlingrabbe.com}
\date{July 1, 2005}

\begin{abstract}
Recently Gupta and Panigrahi have shown how to deterministically identify an unknown Bell state. The
present paper extends their result to deterministic cloning.
\end{abstract}

\keywords{quantum computation, state discrimination, quantum cloning}

\maketitle

Gupta and Panigrahi \cite{MGPP} have recently shown how to deterministically identify an unknown Bell
state. Their procedure involves the use of two ancillary qubits $|a_1\rangle, |a_2\rangle$,
which are transformed to one of the two-qubit states $|00\rangle$, $|01\rangle$,
$|10\rangle$, or $|11\rangle$.  The ancillary qubits are then measured, and depending on the
results of the measurement, the unknown Bell state is identified.

What is easily overlooked here is that the identifying ancillary state $|a_1 a_2\rangle$ is simply the computational basis state from which the associated Bell state may be derived.  This oversight results from the way in which  Bell states are normally designated. So, instead, since Bell states may be derived from the computational basis, let's label them, as in \cite{JOG}, with subscripts  denoting their `origin'.  The correspondence is $0 \leftrightarrow |00\rangle$, 
$1 \leftrightarrow |01\rangle$, $2 \leftrightarrow |10\rangle$, $3 \leftrightarrow |11\rangle$:
\begin{eqnarray}
|b_0\rangle = \frac{1}{\sqrt 2}(|00\rangle + |11\rangle)\\
|b_1\rangle = \frac{1}{\sqrt 2}(|01\rangle + |10\rangle)\\
|b_2\rangle = \frac{1}{\sqrt 2}(|00\rangle - |11\rangle)\\
|b_3\rangle = \frac{1}{\sqrt 2}(|01\rangle - |10\rangle) .
\end{eqnarray}
The transformation to produce the Bell states from the computational basis uses a combination of a Hadamard transformation $H$ and a c-NOT ($\neg$) gate. First apply the Hadamard transform to the left-most qubit. Then apply c-NOT with the left qubit as the source and the right qubit as the target. Shorthand for this transformation is $ \neg (H \otimes \mathbf 1)$, where $\mathbf 1$ is the identity matrix:
\begin{eqnarray}
\neg (H \otimes \mathbf 1) |00\rangle \rightarrow \neg \frac{1}{\sqrt 2}(|0\rangle+|1\rangle) |0\rangle \rightarrow \frac{1}{\sqrt 2}(|00\rangle + |11\rangle) = |b_0\rangle\\  
\neg (H \otimes \mathbf 1) |01\rangle \rightarrow \neg \frac{1}{\sqrt 2}(|0\rangle+|1\rangle) |1\rangle \rightarrow \frac{1}{\sqrt 2}(|01\rangle + |10\rangle) = |b_1\rangle\\  
\neg (H \otimes \mathbf 1) |10\rangle \rightarrow \neg \frac{1}{\sqrt 2}(|0\rangle-|1\rangle) |0\rangle \rightarrow \frac{1}{\sqrt 2}(|00\rangle - |11\rangle) = |b_2\rangle\\  
\neg (H \otimes \mathbf 1) |11\rangle \rightarrow \neg \frac{1}{\sqrt 2}(|0\rangle-|1\rangle) |1\rangle \rightarrow \frac{1}{\sqrt 2}(|01\rangle - |10\rangle) = |b_3\rangle .
\end{eqnarray}
This transformation is reversible.

Now let's designate the Gupta-Panigrahi transform as $T_{gp}$. (Their procedure transforms each ancillary qubit
separately, but this doesn't matter; we can consider the change in the ancillary qubits as a single transformation,
since neither ancillary qubit state depends on the other, or on the result of a measurement.) Thus the transformation of the unknown state $|b_i\rangle$ and the auxillary qubits (in initial state $|a_1 a_2\rangle = |00\rangle$) is as follows:
\begin{eqnarray}
T_{gp}(|b_0\rangle \otimes |00\rangle) = |b_0\rangle \otimes |00\rangle\\
T_{gp}(|b_1\rangle \otimes |00\rangle) = |b_1\rangle \otimes |01\rangle\\
T_{gp}(|b_2\rangle \otimes |00\rangle) = |b_2\rangle \otimes |10\rangle\\
T_{gp}(|b_3\rangle \otimes |00\rangle) = |b_3\rangle \otimes |11\rangle .
\end{eqnarray}
Thus a subsequent measurement of the ancillary qubits yields the associated (unknown) Bell state.  The measurement
does not affect the original Bell state, which is now identified.  The point I wish to emphasize is that the
identifying qubits are simply the computational basis state from which the Bell state may be derived.

Thus, deterministic identification also implies deterministic cloning of the unknown Bell state.  If, instead of
measuring the final state of the ancillary qubits, we instead apply the operation $ \neg (H \otimes \mathbf 1)$
to the final (computational) state of the two ancillary qubits, we obtain a clone of the unknown Bell state:
\begin{eqnarray}
(\mathbf 1 \otimes \mathbf 1 \otimes ( \neg (H \otimes \mathbf 1)))
(T_{gp}(|b_0\rangle \otimes |00\rangle)) = |b_0\rangle \otimes |b_0\rangle\\
(\mathbf 1 \otimes \mathbf 1 \otimes ( \neg (H \otimes \mathbf 1)))
(T_{gp}(|b_1\rangle \otimes |00\rangle)) = |b_1\rangle \otimes |b_1\rangle\\
(\mathbf 1 \otimes \mathbf 1 \otimes ( \neg (H \otimes \mathbf 1)))
(T_{gp}(|b_2\rangle \otimes |00\rangle)) = |b_2\rangle \otimes |b_2\rangle\\
(\mathbf 1 \otimes \mathbf 1 \otimes ( \neg (H \otimes \mathbf 1)))
(T_{gp}(|b_3\rangle \otimes |00\rangle)) = |b_3\rangle \otimes |b_3\rangle .
\end{eqnarray}
We have sucessfully cloned the unknown Bell state with a fidelity of 1.  This compares with a maximum fidelity
of $\frac{5}{6}$ for a Universal Cloning Machine \cite{BDEFMS}.

\end{document}